\documentclass[twocolumn,showpacs,amsmath,amssymb,pra]{revtex4}
\usepackage{color}
\usepackage{graphicx}
\newcommand{\xx}{\mathbf{x}}

\newcommand{\rr}{\mathbf{r}}
\newcommand{\rrho}{\boldsymbol{\rho}}
\newcommand{\kk}{\mathbf{k}}
\newcommand{\qq}{\mathbf{q}}

\begin{document}
\title {Roton immiscibility in a two-component dipolar Bose gas}
\author{Ryan M. Wilson$^1$}
\author{Christopher Ticknor$^2$}
\author{John L. Bohn$^1$}
\author{Eddy Timmermans$^2$} 
\affiliation{$^1$JILA, NIST and Department of Physics, University of Colorado, Boulder, Colorado 80309, USA}
\affiliation{$^2$Theoretical Division, Los Alamos National Laboratory, Los Alamos, New Mexico 87545, USA}
\date{\today}
\begin{abstract}
We characterize the immiscibility-miscibility transition (IMT) of a two-component Bose-Einstein condensate (BEC) with dipole-dipole interactions.  In particular, we consider the quasi-two dimensional geometry, where a strong trapping potential admits only zero-point motion in the trap direction, while the atoms are more free to move in the transverse directions.  We employ the Bogoliubov treatment of the two-component system to identify both the well-known long-wavelength IMT in addition to a roton-like IMT, where the transition occurs at finite-wave number and is reminiscent of the roton softening in the single component dipolar BEC.  Additionally, we verify the existence of the roton IMT in the fully trapped, finite systems by direct numerical simulation of the two-component coupled non-local Gross-Pitaevskii equations.
\end{abstract}
\maketitle

\section{Introduction}
\label{sec:intro}

The phenomenon of Bose-Einstein condensation is characterized by the presence of long-range phase coherence.  Interestingly, phase coherence can persist in a two-component Bose-Einstein condensate (BEC), resulting in an overlapping or miscible mixture of distinguishable components.  This miscible state, however, is only stable for a certain range of interaction parameters and densities.  Indeed, the system can be driven to immiscibility by modulating the density of the two-component BEC, by tuning the relative strengths of the interspecies and intraspecies interactions or by altering the geometry of the trap in which the system is held~\cite{Esry97,Timmermans98,Pu98a,Pu98b,Modugno01,Ronen08PRA}.

To date, most of the experimental work on such systems has been performed using alkali atoms that interact predominantly via short-range potentials~\cite{Myatt97,Papp08,Lercher11}.  For such interactions, the transition to immiscibility is characterized by the parameter $\Delta = g_{11}g_{22}/g_{12}^2-1$, where $g_{11}$ and $g_{22}$ are the intraspecies interaction couplings and $g_{12}$ is the interspecies interaction coupling.   Having $\Delta>0$ implies the stability of the miscible state and $\Delta<0$ implies an unstable miscible state.  Thus, the transition to immiscibility can be seen to originate from the competing strengths of the interspecies and intraspecies contact interactions.  The immiscible-miscible transition (IMT) is characterized by $\Delta=0$, where the interspecies and intraspecies interactions are balanced~\cite{Riboli02,Jezek02}.  The presence of an external trapping potential, however, relaxes this criterion as the trap introduces an additional energy cost for the components to spatially separate~\cite{Ho96,Timmermans98}.

Recently, much effort is being directed towards creating quantum degenerate gases of atoms and molecules that possess non-negligible dipole moments and thus can interact via both short-range and dipole-dipole interactions, which are long-range ($\propto 1/r^3$, where $r$ is the distance between the dipoles) and anisotropic in nature.  Already, experimentalists have succeeded in Bose-condensing atomic $^{52}$Cr with a magnetic dipole moment of $d=6\mu_\mathrm{B}$~\cite{Lahaye09,Bismut10}, atomic $^{164}$Dy with a magnetic dipole moment of $d=10 \mu_\mathrm{B}$~\cite{Lu11PRL} and atomic ${}^{168}$Er with a magnetic dipole moment of $d=7 \mu_\mathrm{B}$~\cite{Aikawa12}, where $\mu_\mathrm{B}$ is the Bohr magneton.  Additionally, progress is being made, for example, towards the condensation molecular RbCs~\cite{Debatin11arXiv}, where the dipolar effects should be considerably larger than in the atomic $^{52}$Cr, $^{164}$Dy and $^{168}$Er BECs.  Clear dipolar effects have been observed in the $^{52}$Cr, $^{164}$Dy and $^{168}$Er BECs, though, in spite of relatively weak dipole-dipole interactions~\cite{Koch08,Lahaye08}.

For polarized dipoles, the anisotropic nature of the interaction leads to attraction when the dipoles are aligned head-to-tail (in the direction of polarization) and thus to an energetic instability in the homogeneous dipolar BEC (DBEC)~\cite{Goral02}.  However, confinement in the direction of polarization can significantly stabilize the dipolar system against instabilities due to inelastic collisions~\cite{Ni08} and three-body loss processes~\cite{Ticknor10}.  To this end, the quasi-two-dimensional (q2D) geometry is sought after to suppress the attractive part of the ddi.  Interestingly, a roton-maxon character is predicted to exist in the quasiparticle dispersion of the single-component DBEC in this q2D geometry~\cite{ODell03,Santos03,Fischer06}, similar to that in superfluid $^4$He, though from a different microscopic origin.  In the q2D dipolar BEC, the roton-maxon dispersion can exist in the dilute, uncorrelated state, while it is precisely the correlations that gives rise the the roton in the superfluid $^4$He system~\cite{Schneider71}.

In this article, we consider a two-component BEC with both short-range and dipole-dipole interactions in the q2D geometry.  Such a system can be realized by dual-condensing different atomic or molecular species, or by using different magnetic sublevels of the same atomic species~\cite{Saito09}.  Indeed, we find that roton physics manifests itself in a unique way in this two-component dipolar system, resulting in a first-order phase transition from the miscible to the immiscible state due to unstable roton-like quasiparticle fluctuations of the two-component BEC.  In practice, this corresponds to a critical length scale at the IMT threshold that is set by the roton wavelength, being on the order of the length scale of the tight trapping potential.  This is in stark contrast to the threshold transition length scale of the non-dipolar condensate, being the phonon length scale, which is typically the longest available length scale in the system.  

Whereas other studies have characterized the IMT of dipolar Bose gases via full mean-field simulations in one dimension~\cite{Gligoric10} and Monte-Carlo simulations in two-dimensions~\cite{Jain11}, we perform full mean-field simulations in addition to an analytic Bogoliubov treatment, allowing us to characterize the IMT very efficiently in a large parameter space.  Additionally, we note that other theoretical works on non-dipolar binary BECs have predicted finite wavelength phenomenon regarding, for example, the boundaries of immiscible systems~\cite{Sasaki09,Kadokura11arXiv,Sasaki11,Suzuki10} and quenches deep into immiscible parameter space~\cite{Sabbatini11}.  These phenomena, however, are not roton-like in that the threshold defining the transition parameters is determined by long-wavelength, or phonon-like excitations in each case.  The q2D binary DBEC is unique in that the IMT has finite wavelength character, or roton character, at the threshold, which is tunable as a function of many system parameters, including the tilt of the polarization field.

This paper is organized as follows.  In section~\ref{sec:mf} we introduce the mean-field formalism for the interacting two-component BEC, discussing the three-dimensional homogenous case in section~\ref{sec:3D} and the quasi-two dimensional homogeneous case in section~\ref{sec:q2Dhomo}.  In section~\ref{sec:bogo} we derive the two-component quasiparticle dispersion, which we use to identify both the familiar long-wavelength and the roton IMT in the large parameter space of the system.  We present our results in section~\ref{sec:results}, including explicit solutions to the coupled Gross-Pitaevskii equations, describing the condensate wave functions of the two-component BEC, for free and radially trapped geometries.  We conclude in section~\ref{sec:conc}.

\section{Mean-Field Theory}
\label{sec:mf}

We consider an ultracold, dilute two-component Bose gas in the presence of two-body $s$-wave and dipole-dipole interactions, labeling the components by the indices $j$ and $k$.  The contact interactions between components $j$ and $k$ are characterized by the $s$-wave scattering lengths $a_{jk}$ through the pseudopotential
\begin{equation}
\label{cpot}
V_{jk}^c(\rr-\rr^\prime) = g_{jk} \delta(\rr-\rr^\prime)
\end{equation}
where $g_{jk} = 2 \pi \hbar^2 a_{jk} / M_{jk}$ is the contact interaction coupling matrix, $M_{jk} = M_j M_k / (M_j + M_k)$ is the reduced mass matrix of two-body system and $M_j$ denotes the mass of a boson of component $j$.  Note that the intraspecies (diagonal) reduced mass is just $M_{jj} = M_j/2$.  Here, $a_{12}=a_{21}$ is the interspecies $s$-wave scattering length and $a_{11}$ and $a_{22}$ are the intraspecies $s$-wave scattering lengths.  The dipole-dipole interaction  (ddi) potential for polarized dipoles is given by the potential
\begin{equation}
\label{dpot}
V_{jk}^d(\rr) = d_j d_k \frac{1 - 3 (\hat{d}\cdot \hat{r})^2}{r^3}
\end{equation}
where $d_j$ is the magnitude of the dipole moment of component $j$, $\hat{d}$ is the direction of the dipole polarization (assumed the same for both components) and $\rr$ is the vector separating the two dipoles.  The full two-body interaction potential is given by $V_{jk}(\rr) = V_{jk}^c(\rr) + V_{jk}^d(\rr)$.  We can write the energy functional for the fully condensed system in terms of the condensate order parameters, or wave functions, $\Psi_j(\rr)$,
\begin{eqnarray}
\label{Efull}
E\left[ \left\{ \Psi_j \right\} \right] = \int d\rr \left[  \sum_{j=1,2} \Psi_j^\star(\rr) \hat{h}_j (\rr) \Psi_j(\rr) \right. \nonumber \\
+ \left. \frac{1}{2} \int d\rr^\prime \sum_{j,k = 1,2}  \Psi^\star_j (\rr) \Psi^\star_k(\rr^\prime) V_{jk} (\rr-\rr^\prime) \Psi_k(\rr^\prime)\Psi_j(\rr) \right].
\end{eqnarray}
Here, $\hat{h}_j(\rr)$ is the single particle Hamiltonian
\begin{equation}
\hat{h}_j(\rr) = -\frac{\hbar^2}{2M_j} \nabla^2 + U_j(\rr),
\end{equation}
and $U_j(\rr)$ is the trapping potential of component $j$.  We normalize each $\Psi_j(\rr)$ to $\int d\rr |\Psi_j(\rr)|^2 = N_j$ where $N_j$ is the number of particles in component $j$ and the total number of particles is given by $N=\sum_{j=1,2} N_j$.  In the energy functional~(\ref{Efull}), the sum over $j,k=1,2$ implies summing over all four combinations of these indices taking on these values.  The factor of $1/2$ in front of this sum takes care of the double counting in the intraspecies interaction terms $j=k$ and of the double counting produced by summing both $\{j,k\} = \{1,2\}$ and $\{j,k\} = \{2,1\}$ for the interspecies interactions. 

The coupled of Gross-Pitaevskii equations (GPEs) for this two-component system are derived by requiring that small variations of $E[\{ \Psi_j \}]$ with respect to $\Psi_j$ vanish, giving
\begin{eqnarray}
\label{GPEj}
i\hbar \partial_t \Psi_j(\rr,t) = \left[ -\frac{\hbar^2}{2M_j}\nabla^2 + U_j(\rr) + \sum_{k=1,2} g_{jk} |\Psi_k(\rr,t)|^2  \right. \nonumber \\
\left. + \sum_{k=1,2} \int d\rr^\prime V_{jk}^d(\rr-\rr^\prime) |\Psi_k(\rr^\prime,t)|^2 \right]\Psi_j(\rr,t).
\end{eqnarray}
Here, we have generalized to the time-dependent form of these equations.  The time-independent forms are recovered by asserting the time dependence $\Psi_j(\rr,t) = \Psi_j(\rr)e^{-i\mu_j t}$ where $\mu_j$ is the chemical potential of component $j$.  Solutions $\Psi_j(\rr)$, corresponding to the fully condensed ground state, are found by minimization of the corresponding energy functional~(\ref{Efull}), and the dynamics can be studied by direct numerical integration of~(\ref{GPEj}).  In practice, as we explain in more detail later, we minimize~(\ref{Efull}) by evolving~(\ref{GPEj}) in imaginary time~\cite{Ruprecht95}.

\subsection{Homogenous Three-Dimensional System}
\label{sec:3D}

In the homogeneous three-dimensional (3D) geometry, there is no trapping potential and the \emph{miscible} ground state of the two-component system can be described simply by the condensate order parameters $\Psi_j(\xx) = \sqrt{n^\mathrm{3D}_j}$, where $n^\mathrm{3D}_j$ is the 3D condensate number density of component $j$.  For the non-dipolar case, the IMT of the homogeneous system is characterized simply by the parameter $\Delta$, introduced in section~\ref{sec:intro}.  This characterization can be made more rigorously by a linear stability study of the miscible state.  Such a study is performed within the Bogoliubov approximation, which results in a quasiparticle description of the two-component BEC.  In this case, a pair of quasiparticles describes in-phase and out-of-phase two-component modes.   The out-of-phase modes play an important role in characterizing the stability of the miscible state.  We save a detailed discussion of such a theory for the case of dipolar interactions in the q2D geometry for the following section~\ref{sec:q2Dhomo}, where the physics of the IMT is more complicated than for the non-dipolar case or homogeneous 3D cases.

When one or both of the components in the homogeneous 3D geometry possesses a non-negligible dipole moment, the polarization breaks the angular symmetry of the mean-field interaction potentials.  In the one-component DBEC, this results in a quasiparticle dispersion that is phonon-like (linear in the long-wavelength limit) but anisotropic.  In this case, the speed of sound $c$ depends on the angle $\theta_\qq$ between the wave vector $\qq$ and the dipole moment, $c\rightarrow c(\theta_\qq)$.  Similarly, the dispersions of the two-component DBEC depend on $\theta_{\qq}$ and are characterized by a $\Delta$ parameter that depends on $\theta_\qq$~\cite{Goral02},
\begin{equation}
\label{Deltaq}
\Delta(\theta_\qq) = \frac{ G_{11}(\theta_\qq) G_{22}(\theta_\qq)}{ G_{12}^2(\theta_\qq) }-1,
\end{equation}
which determines the IMT threshold.  If $\Delta(\theta_\qq)<0$ for any angle $\theta_\qq$, the components of the 3D two-component DBEC are immiscible.  In the above equation,  the $G_{jk}(\theta_\qq)$ matrices characterize both the short-range $s$-wave interactions and the ddi between components $j$ and $k$,
\begin{equation}
\label{gjkq}
G_{jk}(\theta_\qq) = g_{jk} \left\{  1+\epsilon^{dd}_{jk} \left[ 3\cos^2{\theta_\qq}-1 \right]  \right\},
\end{equation}
and $\epsilon^{dd}_{jk}$ characterizes the strength of the ddi between components $j$ and $k$, $\epsilon_{jk}^{dd} = b_{jk}/a_{jk}$, where $b_{jk} = 2 M_{jk} d_j d_k / 3 \hbar^2$ is the dipole length characterizing the ddi between components $j$ and $k$.  With $M_1=M_2$, the interspecies dipole length ($j\neq k$) becomes the geometric mean of the intraspecies dipole lengths, $b_{21}=b_{12}=\sqrt{b_{11}b_{22}}$, and is thus uniquely determined by the intraspecies dipole lengths.  These dipole lengths are defined so that a single component polarized DBEC in a homogeneous 3D geometry requires a positive $s$-wave scattering length $a_{jj}>b_{jj}$ in order to energetically stabilize the system.

It is interesting to consider the case where one component possesses a dipole moment and the other is non-dipolar, say, $b_{22}=0$.  In this case, $\Delta(\theta_\qq)$ is always maximized when $\theta_\qq = \pi/2$, corresponding to two-component quasiparticle propagation in the direction perpendicular to the dipolar polarization.  In this system, the IMT threshold, $\Delta(\theta_\qq)=0$, is first crossed in the direction perpendicular to the dipole polarization, $\theta_\qq=\pi/2$.  With the exception of the angular dependence, the stability condition resembles that of the non-dipolar two-component BEC system.

\subsection{Homogeneous Quasi-Two Dimensional System}
\label{sec:q2Dhomo}

Recently, much effort has been directed towards realizing trapping geometries with very strong confinement in one direction.  Indeed, such geometries significantly stabilize dipolar gases that are polarized in the direction of the strong confinement by suppressing the attractive part of their interactions.  When the characteristic interaction lengths of the trapped atoms or molecules are much larger than the harmonic oscillator length, this system is effectively two-dimensional (2D).  However, when the interaction lengths are sufficiently smaller than the harmonic oscillator length, the system develops q2D character where the zero-point motion in the trapped direction is important in characterizing the interactions~\cite{Santos03,Fischer06} and pure condensation occurs at finite, as opposed to zero temperature~\cite{Petrov00}.  A possible realization of this geometry uses a retro-reflected laser to create a one-dimensional optical lattice potential~\cite{Muller11arXiv}.

We model the two-component Bose system in the q2D geometry with the trapping potential potential $U_j(z) = \frac{1}{2} M_j \omega_z^2 (z^2 + \rho^2/\lambda^2)$ where $\lambda = \omega_z / \omega_\rho \gg 1$ is the trap aspect ratio.  When $\lambda\gg 1$ and the interactions are relatively weak compared to the trapping energy $\hbar \omega_z$, we use the single mode approximation (SMA) where the condensate wave functions are assumed to have the separable form
\begin{equation}
\label{q2Dpsi}
\Psi_j(\xx) = \varphi_j (\rrho) \chi_j(z)
\end{equation}
where $\varphi_j(\rrho)$ is the in-plane wave function normalized to $N_j$, $\int d\rrho |\varphi_j(\rrho)|^2 =N_j$, and $\chi_j(z)$ is the axial wave function.  We take $\chi_j(z)$ to be a Gaussian normalized to unity with width $l_j = \sqrt{\hbar/M_j \omega_z}$,
\begin{equation}
\label{chi}
\chi_j(z) = \frac{1}{\sqrt{l_j}\pi^\frac{1}{4}} \exp{\left[ -\frac{z^2}{2l_j^2} \right]}.
\end{equation}
We assume that both components are trapped in a harmonic trap with the same frequency $\omega_z$, but allow for different masses so the axial wave functions $\chi_j(z)$ can have different widths.

Even in highly oblate traps, the separable ansatz~(\ref{q2Dpsi}) is not exact, except in the case of a non-interacting system.  The interactions, even when weak, also drive the axial wave functions $\chi_j(z)$ away from the Gaussian form.  However, the SMA that we use here significantly simplifies the problem at hand, allowing us to explore a larger region in parameter space, and captures the relevant physics of the system~\cite{Wilson11PRA}.  The SMA is particularly beneficial in that it allows us to reduce the problem to a set of 2D equations by \emph{analytically} integrating out the $z$-dependence in the coupled set of GPEs.  For the dipolar mean field terms, this amounts to calculating an effective q2D interaction potential, given by
\begin{equation}
v_{jk}^d(\rrho-\rrho^\prime) = \int dz \int dz^\prime \chi^2_k(z^\prime) V_{jk}^d(\rr-\rr^\prime) \chi^2_j(z).
\end{equation}
We handle this expression by transforming into momentum space, and thus need the Fourier transforms of $V_{jk}^d(\rr)$ and $\chi^2_k(z)$.  Without loss of generality, we consider a polarizing field $\hat{d} = \cos{\alpha} \hat{z} + \sin{\alpha} \hat{x}$, which describes dipoles that are all tilted by an angle $\alpha$ off of the $z$-axis into the $x$-direction.  The momentum space interaction for dipoles with this configuration is given by~\cite{Goral02}
\begin{equation}
\label{Vk3D}
\tilde{V}_{jk}^d(\kk) = \frac{4\pi}{3} d_j d_k \left\{ \frac{\left(k_z \cos{\alpha} + k_x \sin{\alpha}\right)^2}{k^2}-1 \right\}.
\end{equation}
and the transform of the axial density is given by
\begin{equation}
\label{chik}
\mathcal{F}_\mathrm{1D}\left[ \chi^2_j(z) \right] = \exp{\left[ -\frac{1}{4}k_z^2 l_j^2 \right]},
\end{equation}
where $\mathcal{F}_\mathrm{1D}$ is the 1D Fourier transform operator.  Thus, the effective q2D momentum space ddi is given by~\cite{Fischer06,Ticknor11}
\begin{equation}
\tilde{v}_{jk}^d(\kk_\rho) = D_{jk} F\left[ \frac{\kk_\rho l_{jk}}{\sqrt{2}} \right],
\end{equation}
where $l_{jk} = \sqrt{(l_j^2 + l_k^2)/2}$, $D_{jk} = \frac{2\sqrt{2\pi} \hbar^2 b_{jk}}{M_{jk} l_{jk}}$ is the ddi coupling matrix and $F(\qq) = \cos^2{\alpha} F_\perp(\qq) + \sin^2(\alpha) F_\parallel(\qq)$, where
\begin{equation}
\label{Fperp}
F_\perp(\qq) = 2-3\sqrt{\pi} q e^{q^2} \mathrm{erfc}(q)
\end{equation}
and
\begin{equation}
\label{Fpar}
F_\parallel(\qq) = -1+3\sqrt{\pi} \frac{q_x^2}{q}e^{q^2}\mathrm{erfc}(q),
\end{equation}
and $\mathrm{erfc}(q)$ is the complimentary error function of $q$~\cite{Ticknor11}.  This result~(\ref{Fpar}) can be generalized to describe a polarization field that is rotated by an angle $\eta$ off of the $x$-axis, by taking $q_x \rightarrow q_d$, where $q_d = \sqrt{q_x \cos^2{\eta} + q_y \sin^2{\eta}}$ is the wave number in the direction of the polarization tilt.

Similar calculations can be carried out for the kinetic, potential and contact interaction terms to yield a coupled set of GPEs that govern the in-plane wave functions $\varphi_j(\rrho)$ of the q2D system,
\begin{widetext}
\begin{equation}
\label{q2DGPEj}
i \hbar \partial_t \varphi_j(\rrho,t) = \left\{ -\frac{\hbar^2}{2M_j} \nabla_\rho^2  + \frac{1}{2 \lambda^2} M_j \omega_z^2 \rho^2+ \sum_{k=1,2} \left[ \frac{g_{jk}}{\sqrt{2\pi} l_{jk}} |\varphi_k(\rrho,t)|^2 +  \int d\rrho^\prime \, v_{jk}^d(\rrho-\rrho^\prime) |\varphi_k(\rrho^\prime,t)|^2 \right] \right\}\varphi_j(\rrho,t).
\end{equation}
\end{widetext}
We calculate the dipolar interaction in the last term in Eqs.~(\ref{q2DGPEj}) by employing the convolution theorem,
\begin{equation}
\int d\rrho^\prime v_{jk}^d (\rrho-\rrho^\prime) |\varphi_k(\rrho^\prime,t)|^2 = \mathcal{F}_\mathrm{2D}^{-1}\left[ \tilde{v}_{jk}^d(\kk_\rho) \tilde{n}_k(\kk_\rho,t) \right],
\end{equation}
where $\mathcal{F}_\mathrm{2D}$ is the 2D Fourier transform operator and $\tilde{n}_k(\kk_\rho,t) = \mathcal{F}_\mathrm{2D}\left[ |\varphi_k(\rrho,t) |^2 \right]$.  Eqs.~(\ref{q2DGPEj}) fully describe the two-component DBEC in the q2D geometry (in the mean-field framework), where the components can have different masses and interaction character.  It is interesting to note that the off-diagonal elements of the ddi coupling matrix $D_{jk}$ are determined uniquely by its diagonal elements while the off-diagonal elements of the $s$-wave contact interaction coupling matrix $g_{jk}$ are, in principle, independent of the diagonal elements.  Physically, the latter are determined by the microscopic structure of the components and are tunable via magnetic Fano-Feshbach resonance~\cite{Regal03,Werner05}.

\subsection{Linear Stability:  Bogoliubov Theory}
\label{sec:bogo}
 
 An instructive case to consider is the pure q2D case, or the case with no radial trapping potential ($\lambda \rightarrow \infty$) so the system is homogeneous in the $x$-$y$ plane.  In the miscible state away from any instabilities, the condensate wave functions can be written as $\varphi_j(\rrho,t) = \sqrt{n_j} e^{-i\mu_j t}$ where $n_j$ is the integrated 2D density of the $j^\mathrm{th}$ component.  In the immiscible state, however, the continuous translational symmetry of the system is broken.  We can study this immiscibility-miscibility transition (IMT) by considering small deviations from the ground-condensed miscible states in the form of Bogoliubov quasiparticles, $\varphi_j(\rrho,t) \rightarrow \sqrt{n}_j \left( 1 + \delta \psi_j(\rrho,t) \right)e^{-i\mu_j t}$, where $\delta \ll 1$ and
 \begin{equation}
 \psi_j(\rrho,t) = u_j e^{i\qq_\rho\cdot \rrho}e^{-i \omega t} + v^\star_j e^{-i\qq_\rho\cdot \rrho}e^{i \omega t} 
 \end{equation}
where $u_j$ and $v_j$ are the Bogoliubov particle and hole amplitudes, respectively, and obey the normalization $|u_j|^2 - |v_j|^2 = 1$~\cite{Dalfovo99}.  We derive a set of equations for the frequencies $\omega$, the Bogoliubov de Gennes (BdG) equations, by linearizing the Bogoliubov ansatz about the small parameter $\delta$ in the GPEs~(\ref{q2DGPEj}).  The same equations can be derived from a second quantized theory, by diagonalizing the full two-component Hamiltonian in the Bogoliubov approximation~\cite{Tommasini03}.  These equations can be written in matrix form as~\cite{Sun10}
\begin{equation}
\label{BdG}
\mathcal{H}\mathbf{u} = \omega \mathbf{u}
\end{equation}
where $\mathbf{u}^\mathrm{T} = [u_1,v_1,u_2,v_2]$ and the two-component BdG Hamiltonian $\mathcal{H}$ is given by the $4\times 4$ matrix
\begin{equation}
\label{BdGH}
\mathcal{H} = 
\left( \begin{array}{cc}
\mathcal{B}_{11} & \mathcal{B}_{12} \\
\mathcal{B}_{21} & \mathcal{B}_{22} \\
\end{array}\right)
\end{equation}
and the $\mathcal{B}_{jk}$ are $2\times 2$ sub-matrices, and are given by
\begin{equation}
\label{BdGH}
\mathcal{B}_{jk} = 
\left( \begin{array}{cc}
\frac{\hbar^2 q^2}{2M_j} \delta_{jk} + h_{jk} & h_{jk} \\
-h_{jk} & -\frac{\hbar^2 q^2}{2M_j} \delta_{jk} - h_{jk}
\end{array}\right)
\end{equation}
where $h_{jk} = h_{jk}(\qq)$ is a function of the quasiparticle momentum,
\begin{equation}
\label{hjk}
h_{jk}(\qq) = n_k\left( \frac{g_{jk}}{\sqrt{2\pi}l_{jk}} + D_{jk} F\left[ \frac{\qq l_{jk}}{\sqrt{2}} \right] \right).
\end{equation}
An algebraic diagonalization of the BdG equations~(\ref{BdG}) yields the two-branch dispersion of the miscible q2D system,
\begin{widetext}
\begin{equation}
\label{disp}
\omega_{\pm}^2(\qq) = \frac{1}{2} \left\{ \omega_1^2(\qq) + \omega_2^2(\qq) \pm \sqrt{ \left( \omega_1^2(\qq)-\omega_2^2(\qq)  \right)^2  + \frac{4 q^4}{M_1 M_2} h_{12}(\qq) h_{21}(\qq)   } \right\}
\end{equation}
\end{widetext}
where $\omega_j(\qq)$ are the single component Bogoliubov dispersions
\begin{equation}
\label{bogo1}
\omega_j^2(\qq) = \frac{q^2}{2M_j} \left( \frac{q^2}{2M_j} + 2 h_{jj}(\qq) \right).
\end{equation}
The two-component dispersion~(\ref{disp}) is identical to that of the homogeneous system with contact interactions~\cite{Tommasini03} but with full momentum-dependent interaction coupling.  We solve for the BdG eigenvectors $\mathbf{u}$ numerically.  From these eigenvector solutions, we identify the upper ($+$) and lower ($-$) branches of the two-component dispersion~(\ref{disp}) as corresponding to in-phase and out-of-phase modes, respectively.  As is clear from the form of the dispersion, $\omega_-$ will always be lower in energy than $\omega_+$ and is thus the relevant branch regarding stability of the miscible state.  In this work, we consider only positive intraspecies and interspecies $s$-wave scattering lengths.  In this case, and in the absence of the ddi, the single components are always stable and any instability, corresponding to $\mathrm{Im}[\omega_-]\neq0$, signifies a transition to an immiscible state.  When the ddi is present, however, the single particle dispersions can present dynamical roton instabilities when the effective dipole length is sufficiently larger than the positive $s$-wave scattering length, due to the momentum-dependence of the ddi in the q2D geometry~\cite{Santos03,Fischer06}.  As a result, imaginary parts of the lower branch of the two-component dispersion could correspond to transitions to immiscible states or to collapse of the miscible system, much like the roton collapse seen in the single-component DBEC.

Before proceeding, we consider the most radical deviation from the simple Gaussian $z$-dependence expressed by Eq.~(\ref{chi}), being phase separation in the trap ($z$) direction.  In the absence of the trap, miscibility in this ($z$) direction corresponds to stable two-component quasiparticle propagation in this direction, which corresponds to stable quasiparticle propagation at an angle $\alpha$ off of the $z$-axis, where $\alpha$ is the polarization angle, for the homogeneous 3D case discussed in section~\ref{sec:3D}.  Thus, for the system to be miscible in the trap direction, the parameters must satisfy (from Eq.~(\ref{Deltaq}))
\begin{equation}
\label{q2Dmisccrit}
\Delta(\alpha)>0.
\end{equation}
Indeed, all of the examples we consider here satisfy this criterion, so the assumption that there is no spatial separation or immiscibility in the $z$ direction is a good one.  Even if Eq.~(\ref{q2Dmisccrit}) is not satisfied, however, the system is likely still miscible in the trap direction due to the tendency of the strong trapping potential to force a system to miscibility.  In this case, the criteria for miscibility in the trap direction in the homogeneous q2D geometry follows directly from previous studies of the IMT in a two-component BEC in a trap, as the ddi can be treated by using a modified $s$-wave scattering length in this case~\cite{Santos03}.  This was recently considered for specific dipolar species in~\cite{Xi11}.

\section{Results}
\label{sec:results}

For simplicity, we proceed by considering a two-component system with equal integrated densities $n_1=n_2=n$ and masses $M_1=M_2=M$, which we refer to as the ``balanced'' system.  Having such a balanced system results in the two components sharing an axial wave function, $\chi_1(z)=\chi_2(z)$.  In the homogeneous 3D geometry, the stability of the mixture does not depend on the direction of the dipole moment.  Such is not the case, however, in the q2D geometry.  If the dipoles are polarized in-plane, corresponding to $\alpha=\pi/2$, the system acquires the same energetic stability criterion as the homogeneous 3D case.  If the dipoles are polarized in the trap direction, corresponding to $\alpha=0$, the criterion for  energetic instability becomes $a_{jj} > -2b_{jj}$~\cite{Santos03}.  For now, we consider the $\alpha=0$ case.  Additionally, from here forward we set $n l_z^2 = 1$, so that the interactions are characterized solely by the ratio of the interaction lengths to the axial harmonic oscillator length $l_z$.

\subsection{Roton Immiscibility for $\alpha=0$}

For the short-range $s$-wave interactions, we consider scattering lengths that would result in a slightly immiscible system in the absence of the ddi.  For now, we take $a_{11}/ l_z = a_{22} / l_z = 1.0$ and $ a_{12} / l_z = 1.05$.  Additionally, we fix $b_{11} / l_z = 2/3$ and explore the stability of the miscible system as a function of $b_{22} / l_z$ using the Bogoliubov theory laid out in the previous section.

\begin{figure}[t]
\includegraphics[width=.95\columnwidth]{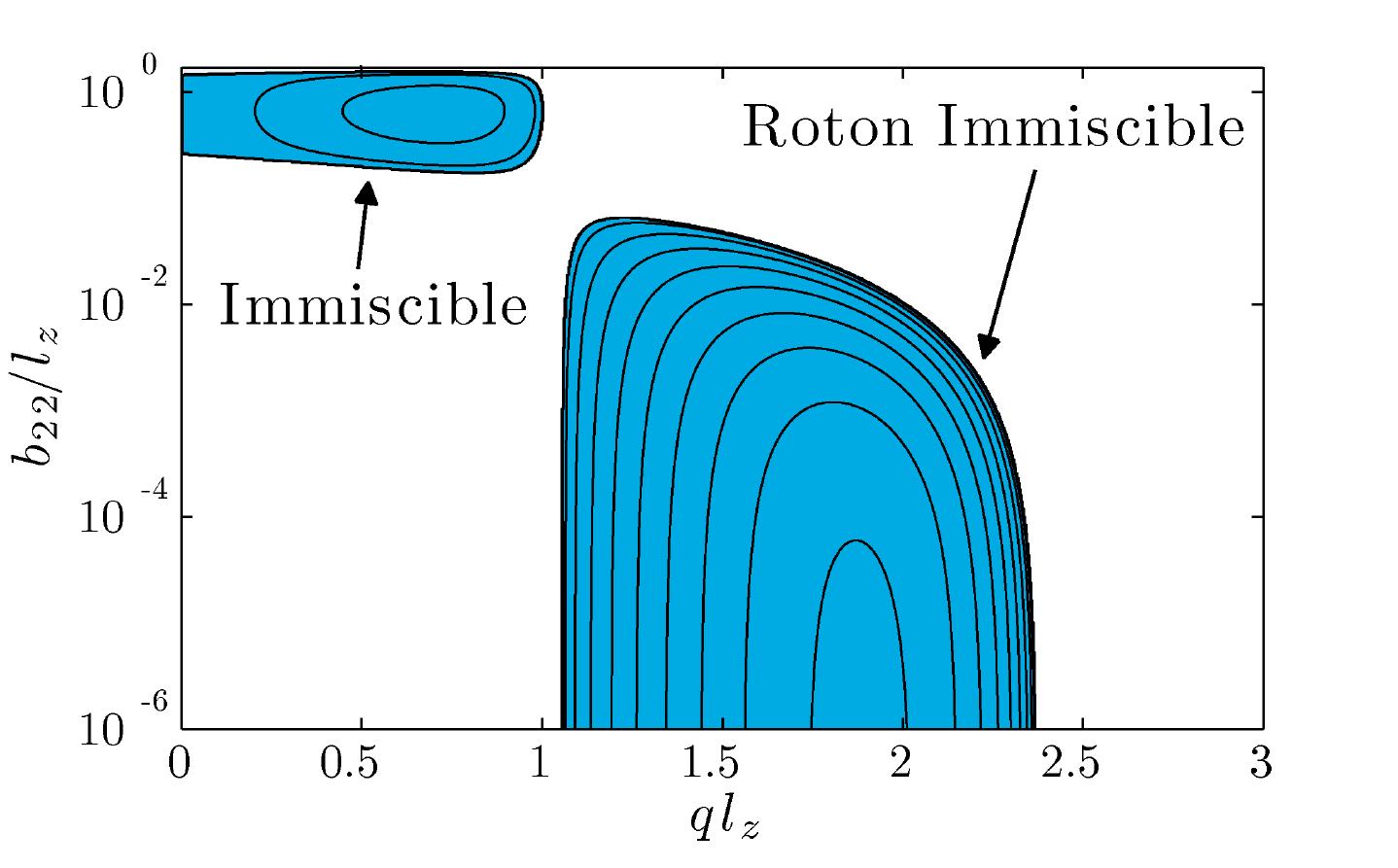}
\caption{\label{fig:stab} (color online) The phase diagram of the balanced two-component system in the q2D geometry with dipole lengths $b_{11}/l_z = 2/3$ and scattering lengths $a_{11}=a_{22}=1$ and $a_{12}=1.05$ as a function of the dipole length of component-2, $b_{22}/l_z$ and quasiparticle wave number $q l_z$.  The familiar long-wavelength immiscible region and the roton-immiscible region are labeled and colored, and the white region signifies the existence of a stable miscible state.  The roton-immiscible region extends down to $b_{22}/l_z=0$.  The contours in the shaded region mark increments of $0.1 \omega_z$.}
\end{figure}

It is straightforward to predict the behavior of the two-component system when $b_{11}=b_{22}$, that is, when both components are equally dipolar.  In this case, the intra- and inter-species dipolar interactions are equally repulsive and their effects \emph{cancel} in the dispersion $\omega_-(\qq)$.  As a consequence, the ddi plays no role in the IMT of the system.  The IMT is instead solely determined by the short-range parameters.  Indeed, this is seen in figure~\ref{fig:stab}, where the imaginary part of the lower branch of the two-component Bogoliubov dispersion, $\omega_-(\qq)$ from Eq.~(\ref{disp}), is shown as as a function of the wave number $q l_z$.  For a range of $b_{22} / l_z \simeq b_{11}/l_z =  2/3$, there is a long-wavelength immiscibility, labeled and shown by the shaded region in this figure.  For $b_{22} \neq b_{11}$, however, the difference in the dipole moments of the two components plays an important role in characterizing the stability of the miscible state.  For example, reducing $b_{22}$ relative to $b_{11}$ stabilizes the long-wavelength instability of the miscible state, as seen in the stable gap for a range of $b_{22}/l_z$ in figure~\ref{fig:stab}.

Another important example of this, and indeed a key result in this paper, occurs for smaller values of $b_{22}$.  For the parameters given above, and for $b_{22}/l_z \lesssim 0.1$, there exists another region where $\mathrm{Im}[\omega_-(\qq)] \neq 0$.  This region occurs at finite, nonzero wave number, and is characterized by the softening of a roton-like feature in the $\omega_-$ quasiparticle dispersion.  As in the case of a single component DBEC, the softening of the roton dispersion signals an instability of the q2D-homogeneous system.  Unlike the single component DBEC, however, the instability does not lead to collapse but instead results in an immiscible density pattern.Ê The features in the density pattern have a length scale that is the inverse of the momentum at which the the roton dispersion touches the axis.Ê We term this kind of phase-separation `roton immiscibility.'  While the case $b_{22}/l_z=0$ is not shown in this figure (due to the logarithmic scaling), the region of roton immiscibility extends down to this limit.

\begin{figure}[t]
\includegraphics[width=.95\columnwidth]{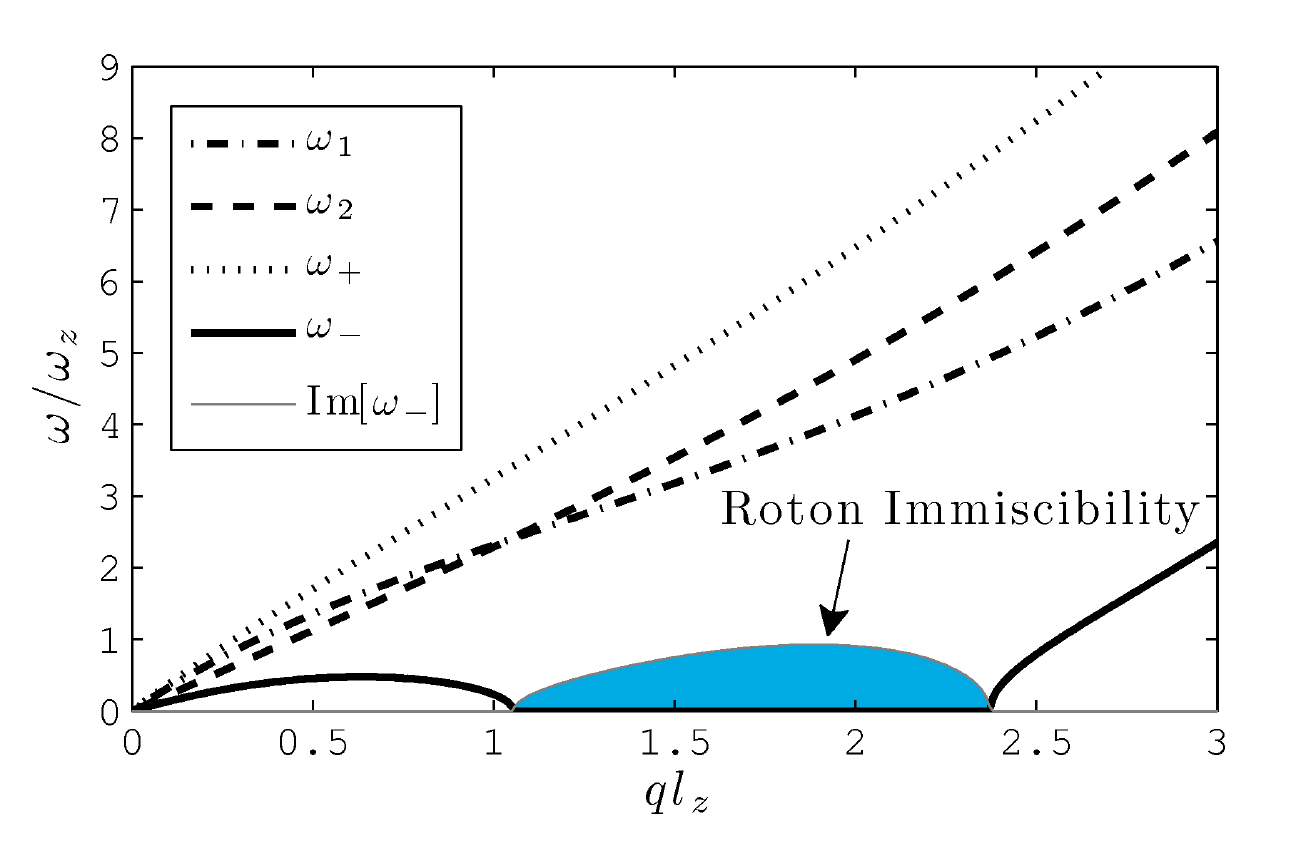}
\caption{\label{fig:disp} (color online) Dispersions of the balanced, miscible, two-component q2D DBEC with $a_{11}/l_z = a_{22}/l_z = 1$, $a_{12}/l_z = 1.05$, $b_{11}/l_z=2/3$ and $b_{22}/l_z = 0$.   The lower branch of the two-component dispersion, $\omega_-(\qq)$, corresponding to out-of-phase quasiparticle excitations, exhibits a roton-like instability corresponding to a transition to immiscibility.  The imaginary part of the dispersion is shaded in teal.  }
\end{figure}

The limit of $b_{22}/l_z=0$ is, in fact, useful in revealing the nature of the roton immiscibility.  In this case, the only dipolar interactions in the system occur within component 1.  The dispersion of component 1, $\omega_1(\qq)$, does not possess a roton minimum, but is not purely phonon-like, either, due to the momentum dependence of the interaction $h_{11}(\qq)$, Eq.~(\ref{hjk}).  This dispersion is shown by the dash-dotted line in figure~\ref{fig:disp}, along with the dispersion of component 2, $\omega_2(\qq)$ (dashed line), which is purely phonon-like due to the purely short-range nature of these interactions.  The $a_{12}/l_z$ ratio that characterizes the interspecies repulsion is then sufficiently strong to drive the system to immiscibility, but on a length scale set by the emerging roton in component 1.  The lower branch of the two-component dispersion, $\omega_-(\qq)$, is shown in figure~\ref{fig:disp}, as well, where its imaginary part, signifying the transition to immiscibility, is labeled and shaded.  

\begin{figure}[t]
\includegraphics[width=1\columnwidth]{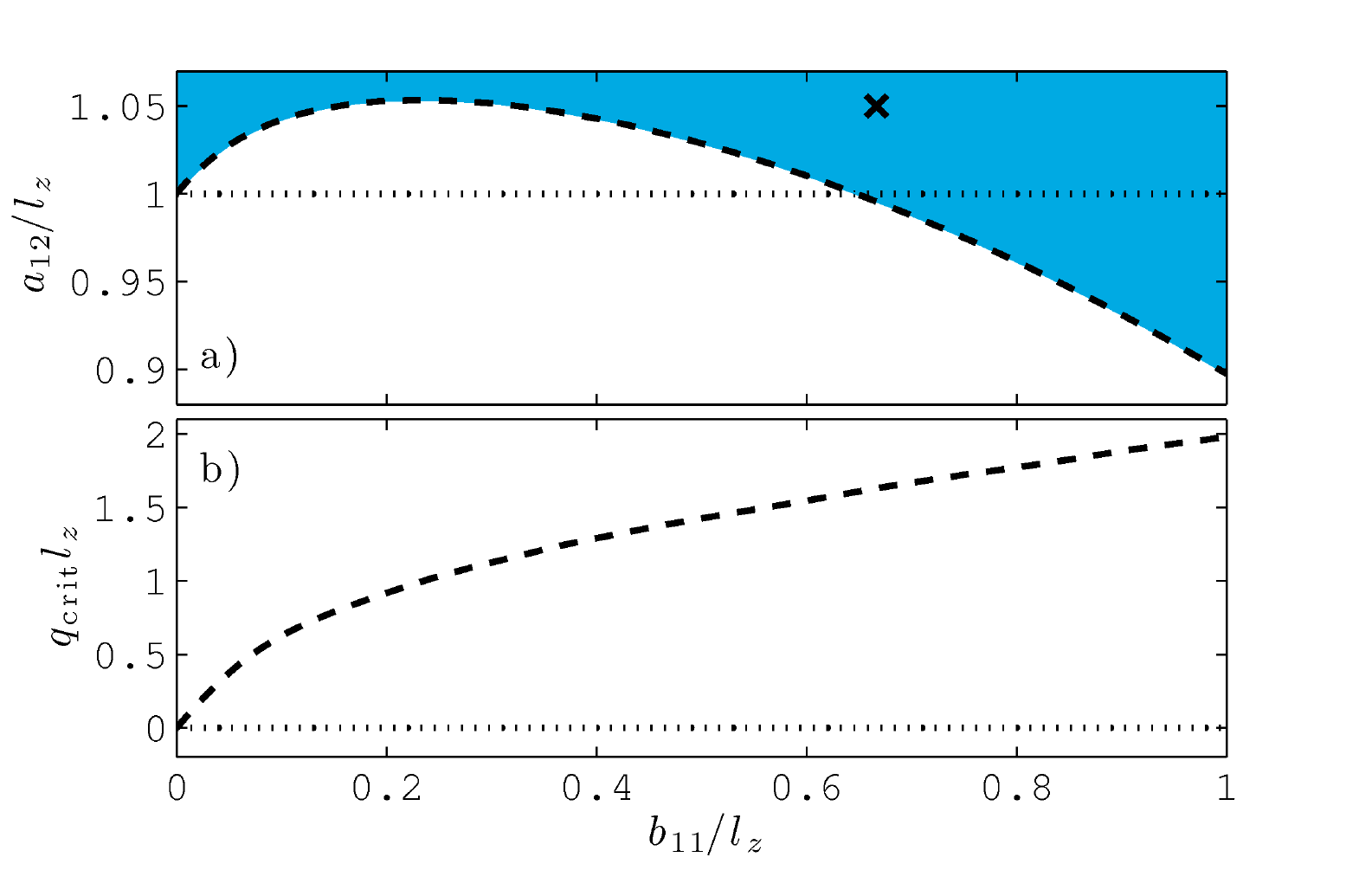}
\caption{\label{fig:para} (color online) Critical interspecies $s$-wave scattering length $a_{12}/l_z$ (top panel) and wave number (bottom panel) for the balanced two-component q2D DBEC as a function of the dipole length of component 1, $b_{11}/l_z$, with $a_{11}/l_z = a_{22}/l_z = 1$ and $b_{22}/l_z = 0$.  The parameters for which the system is immiscible are shaded teal in the top panel.  The non-dipolar IMT is seen for $b_{11}/l_z =0$, where the IMT threshold occurs at $a_{12}/l_z = 1$ ($\Delta=0$) and $q_\mathrm{crit} l_z = 0$.  The roton immiscibility emerges for nonzero $b_{11}/l_z$, and decreases in characteristic wavelength as $b_{11}/l_z$ is increased.  The black ``x'' in the top panel marks $b_{11}/l_z = 2/3$ and $a_{12}/l_z = 1.05$, the parameters for the example roton IMT used in figures~\ref{fig:disp} and~\ref{fig:rotonim}.}
\end{figure}

We map the roton immiscibility phase boundary for the balanced system with $a_{11}/l_z = a_{22}/l_z =1$ and $b_{22}/l_z=0$ in figure~\ref{fig:para}(a).  This figure shows the IMT threshold as a function of the interspecies scattering length $a_{12}/l_z$ and the dipole length of component 1, $b_{11}/l_z$.  Interestingly, the roton immiscibility persists for all non-zero values of $b_{11}$ when the intraspecies scattering length is larger than a critical interspecies scattering length.  This is shown in figure~\ref{fig:para}(a), where we plot the threshold intraspecies scattering length as a function of $b_{11}/l_z$, above which the system is immiscible (shaded) and below which the system is miscible.  Interestingly, for $b_{11}/l_z \gtrsim 0.6$ the threshold for roton immiscibility occurs at $a_{12} / l_z < 1$.  In the absence of any ddi, the transition to miscibility occurs when $a_{12}/l_z \geq 1$, which is shown by the black dotted line in this figure.

The onset of instability at a finite momentum is a roton-specific feature that signals a first-order (zero-temperature) phase transition.  The inverse of the critical momentum $q_\mathrm{crit}$ at which the roton in the $\omega_-$ dispersion softens indicates the length scale at which immiscible (single condensate) density features nucleate if the system evolves from a homogeneous ground state.  Figure~\ref{fig:para}(b) shows the critical wave number $q_\mathrm{crit} l_z$ at which the dispersion $\omega_-(\qq)$ first develops a non-zero imaginary part with increasing $a_{12}/l_z$ as a function of $b_{11}/l_z$. For $b_{11}/l_z = 0$, corresponding to a completely non-dipolar system, the immiscibility transition occurs at $q_\mathrm{crit} l_z = 0$, which is  the familiar long-wavelength immiscibility found in BECs with only short-range interactions.  As $b_{11}/l_z$ is increased, however, the critical wave number for immiscibility increases, corresponding to a transition to immiscibility with roton, or finite wavelength character.  Indeed, when $b_{11} /l_z = 1$, the transition to immiscibility occurs at the large wave number $q l_z \sim 2$, corresponding to a transition wavelength $\lambda_\mathrm{crit} \sim \pi l_z$.  

We proceed to further identify the roton instability of the miscible state as a transition to immiscibility, and not an instability to a \emph{collapsed} state, by directly solving the coupled GPEs, Eqs.~(\ref{q2DGPEj}), in the absence of a radial trap, corresponding to $\lambda \rightarrow \infty$.  To find the stationary ground state, we sample the condensate wave functions of the two components on a numeric grid and employ the imaginary-time evolution algorithm, stopping when the total energy is converged to a part in $10^{8}$.  We choose a grid of size $512\times 512$ with the spatial extent $x \in [-x_\mathrm{max},x_\mathrm{max}]$ and $y \in [-y_\mathrm{max},y_\mathrm{max}]$.  These real-space limits $x_\mathrm{max}$ and $y_\mathrm{max}$ are chosen so that the spatial resolution of the numeric grid is sufficiently smaller than both the healing length of the system and the axial harmonic oscillator length $l_z$.  To initiate the algorithm, we break the symmetry of the system by seeding the initial homogeneous guesses for the condensate wave functions with numeric noise in the form of two-component quasiparticles.  That is, we take for component $j$~\cite{Gardiner99}
\begin{eqnarray}
\label{imagtimeseed}
\varphi_j(\rrho,t=0) &\rightarrow& \sqrt{n} \left\{ 1 + \sum_l \sqrt{\nu_{l}} \, e^{2\pi i \alpha_l} \right.\nonumber \\
&\times& \left. \left[ u_{j,l} e^{-i \qq_{l} \cdot \rrho} + v^\star_{j,l}e^{i \qq_{l} \cdot \rrho} \right] \right\}
\end{eqnarray}
where $\{ \alpha_l\}$ are random numbers such that $\alpha_l \in [0,1]$ for all $l$ and $\nu_{l} = N_{l}/N_{0}$, $N_{0} = n l_z^2$ and $N_{l}$ is the number occupation of the quasiparticle state of component $j$ with energy $\omega_{-,l}$, given by the Bose-Einstein distribution,
\begin{equation}
\label{BEdist}
N_{l} = \left\{ e^{\omega_{-,l}/k_B T}-1 \right\}^{-1},
\end{equation}
where $T$ is the temperature of the system.  While we choose $T=100 \, \mathrm{nK}$, this does not carry strong physical meaning when evolving Eqs.~(\ref{q2DGPEj}) in imaginary time, as this equation is dissipative.  We calculate the $u_{j,l}$ and $v^\star_{j,l}$ quasiparticle amplitudes via numeric diagonalization of~(\ref{BdGH}).  Additionally, we choose $\qq_{l}$ such that an integer number of quasiparticle wavelengths fit in the grid, so that the periodic boundary conditions of the system are satisfied.  Such a restriction, however, is not important when evolving the system in imaginary time, as any unphysical, high-momentum components of the initial wave functions that are rooted in relaxing this restriction will quickly dissipate.  Indeed, we choose a random direction for each $\qq_{l}$ to introduce noise that does not share the symmetry of the numeric grid.  Finally, we sample $\omega_{-,l}$ from the dispersion $\omega_-(\qq_l)$, noting that these energies are the same for both components.  

\begin{figure}[t]
\includegraphics[width=1\columnwidth]{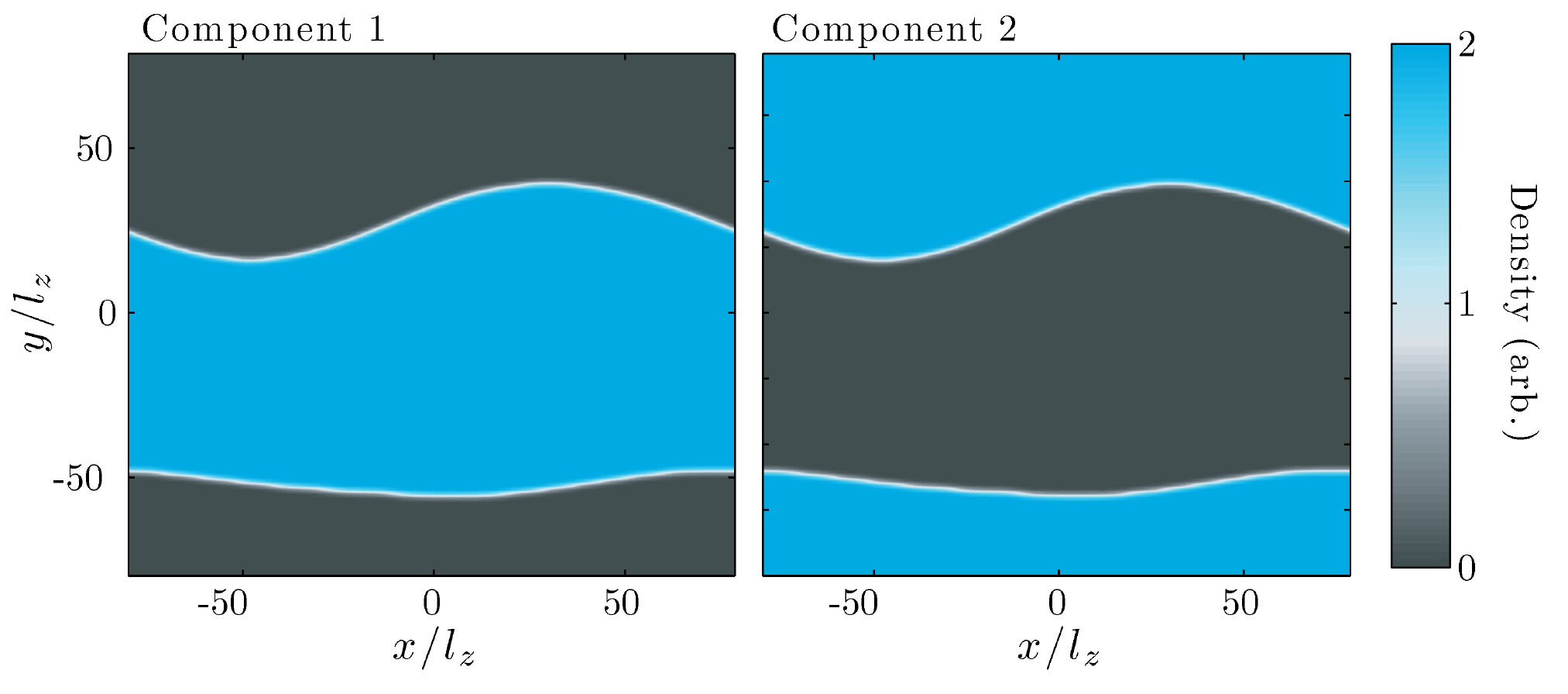}
\caption{\label{fig:regim} (color online) Densities $n_1(\rrho)=|\varphi_1(\rrho)|^2$ (left panel) and $n_2(\rrho) = |\varphi_2(\rrho)|^2$ (right panel) corresponding to stationary solutions of the coupled GPEs~(\ref{q2DGPEj}) (converged in energy to a part in $10^{8}$ in imaginary time evolution) for the non-dipolar balanced system, with $a_{11}/l_z = a_{22}/l_z = 1$, $a_{12}/l_z = 1.05$ and $b_{11}/l_z = b_{22}/l_z=0$.  Here, the immiscibility is phonon-like and occurs on the longest length scale available in the system.}
\vspace{8pt}
\includegraphics[width=1\columnwidth]{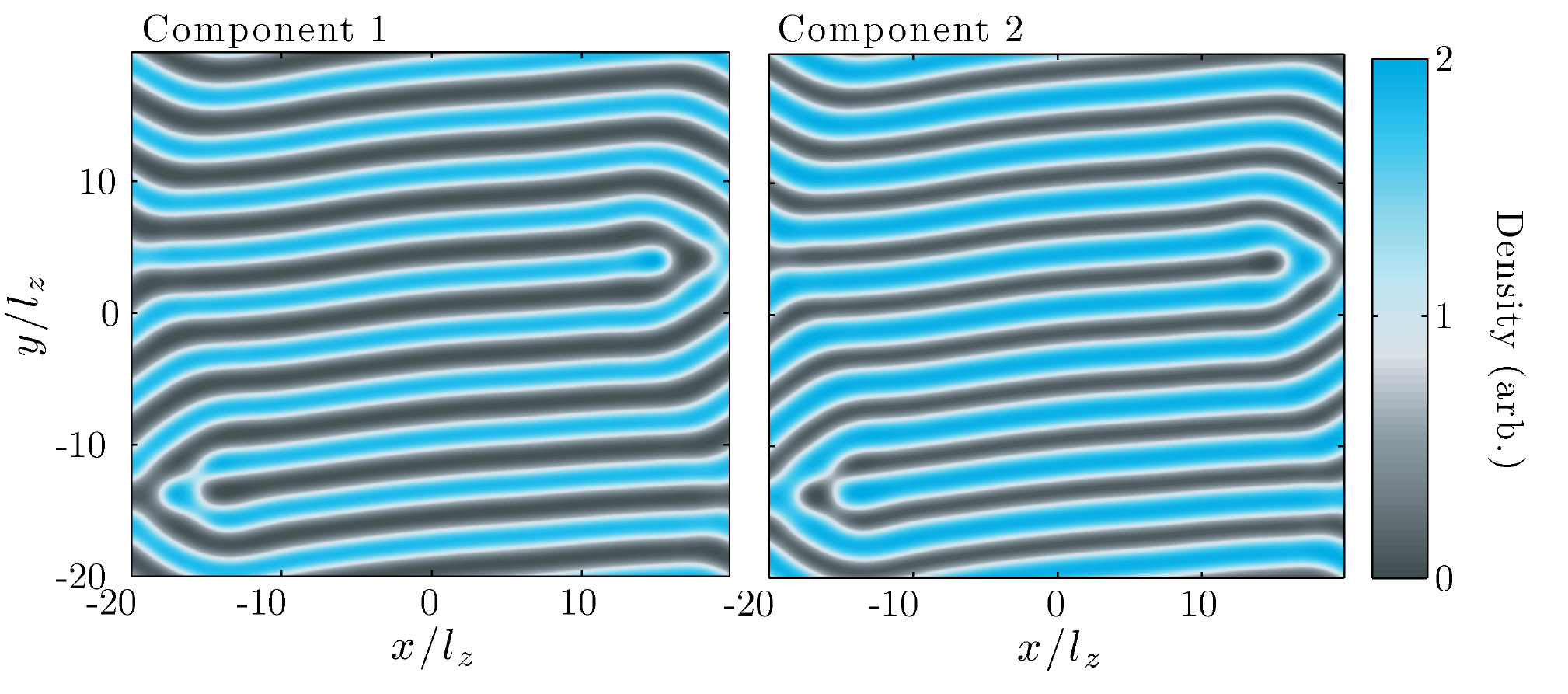}
\caption{\label{fig:rotonim} (color online) Densities $n_1(\rrho)=|\varphi_1(\rrho)|^2$ (left panel) and $n_2(\rrho) = |\varphi_2(\rrho)|^2$ (right panel) corresponding to stationary solutions of the coupled GPEs~(\ref{q2DGPEj}) (converged in energy to a part in $10^{8}$ in imaginary time evolution) for the balanced system, with $a_{11}/l_z = a_{22}/l_z = 1$, $a_{12}/l_z = 1.05$, $b_{22}/l_z=0$ and $b_{11}/l_z = 2/3$.  Here, the immiscibility is roton-like and occurs on the length scale $\sim 2\pi l_z$.  }
\end{figure}

Using this algorithm to find stationary solutions of the coupled GPEs, Eqs.~(\ref{q2DGPEj}), we first explore the case when $b_{11}/l_z = b_{22}/l_z = 2/3$.  As mentioned above, this case is immiscible but the dispersion is phonon-like and does not result from the softening of the roton dispersion feature.  In accordance, the ground state solution to the coupled GPEs, shown in figure~\ref{fig:regim}, exhibits immiscibility at the longest available wavelength.

The case where $b_{22}/l_z = 0$ and $b_{11}/l_z = 2/3$, however, is in stark contrast to the case where both components possess equal dipole moments.  The two-component Bogoliubov theory predicts a roton instability of the miscible mixture in the $\omega_-(\qq)$ dispersion.  Indeed, we find that the ground state of this system exhibits immiscibility on a much shorter length scale, being the length scale of the roton.  The condensate densities of the two components are shown for this case in figure~\ref{fig:rotonim}.

For the cases shown in figures~\ref{fig:regim} and~\ref{fig:rotonim}, the ground state solutions depend on the numeric noise, in this case the occupation of single quasiparticles with well-defined, albeit random, standing wave orientations, that is used to seed the condensate at the beginning of the imaginary time evolution.  For example, if wave vectors are chosen such that $\qq = q \hat{x}$, we find that the immiscible ground states possess density fluctuations only in the $x$-direction.  Similarly, we performed simulations with random noise in the form of small-amplitude random numbers sampled at each grid point.  The immiscible ground state for the case with one dipolar and one non-dipolar component (the roton immiscible ground state) shows a speckle, or bubble-like pattern, fluctuating on a length scale $\sim 4 l_z$.  In the cases discussed here, the origin of the sensitivity to the symmetry of the initial seeding of the wave functions is the fact that there is no internal system bias for the direction of the immiscibility ($\alpha=0$).  However, the anisotropy of the ddi can be exploited to introduce an anisotropic momentum-dependence in the system interactions ($\alpha\neq 0$), thus breaking the azimuthal symmetry of the system and creating a directional bias.

\subsection{Roton Immiscibility for $\alpha > 0$}

The behavior of the quasiparticle dispersion of the single-component q2D DBEC was found to exhibit interesting quasiparticle dispersion character as a function of polarization direction~\cite{Ticknor11}.  More specifically, an anisotropic roton emerges as a function of polarization angle $\alpha$ for certain interaction strengths and densities, at wave vectors perpendicular to the direction of the polarization tilt.  Here, we find an analogous phenomenon in the two-component system.

\begin{figure}[t]
\includegraphics[width=.95\columnwidth]{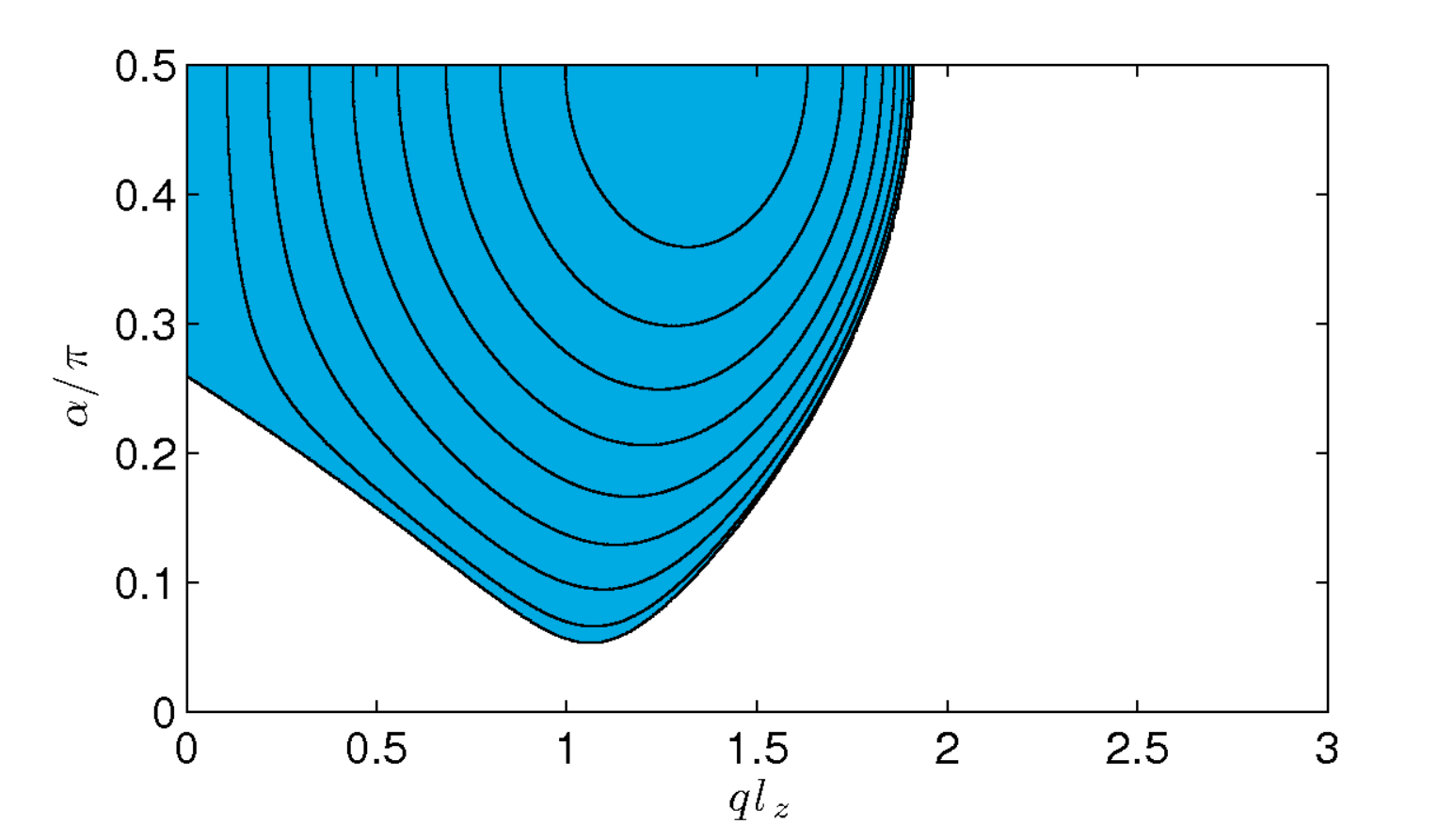}
\caption{\label{fig:stabang} (color online) The shaded region shows the non-zero imaginary part of the two-component dispersion $\omega_-(\qq)$ in the direction perpendicular to the polarization tilt for $a_{11}/l_z = a_{22}/l_z = 1$, $a_{12}/l_z = 1.05$, $b_{11}/l_z = 1/4$ and $b_{22}/l_z = 0$ as a function of polarization angle $\alpha$ and quasiparticle wave number $q l_z$.  The miscibility of the system for $\alpha=0$ is reflected in the purely real character of the dispersion at this polarization angle, and the roton-immiscibility is seen as the emergence of a non-zero imaginary part of $\omega_-(\qq)$ at finite wave number $ql_z \sim 1$ at $\alpha / \pi \sim 0.05$.  The contours in the shaded region mark increments of $0.1 \omega_z$.}
\end{figure}

For now, consider a two-component BEC with the parameters used above, but with a smaller dipole length in component 1, $b_{11}/l_z = 1/4$, and no dipolar character in component 2, $b_{22}/l_z = 0$.  If the polarization axis is perpendicular to the trapping direction, corresponding to $\alpha=0$, the ground state of this system is a miscible mixture of the two components.  As the polarization axis is tilted, however, an instability to a miscible state emerges as the character of the ddi in component 1 becomes more attractive in this direction.  We plot the imaginary part of the lower branch of the two-component dispersion relation, $\omega_-(q)$, for the component of the quasiparticle wave vector that is perpendicular to the polarization tilt as a function of the tilt angle $\alpha$ and the quasiparticle momentum in figure~\ref{fig:stabang}.  The miscible ground state is reflected in the purely real dispersion at $\alpha=0$.  A non-zero imaginary part emerges at $\alpha / \pi \sim 0.05$ at finite, non-zero wave number, corresponding to a roton immiscibility that is emergent with polarization tilt.  This roton immiscibility has the same origin as that discussed earlier, being the momentum dependence of the ddi in the q2D geometry, though this momentum dependence is anisotropic and possesses angular dependence for $\alpha \neq 0$.  Indeed, no immiscibilities are predicted \emph{in} the direction of the polarization tilt for any $\alpha$, suggesting that the striped structure of the immiscible state can be controlled by the proper adjustment of the polarization field.

\begin{figure}[t]
\includegraphics[width=.95\columnwidth]{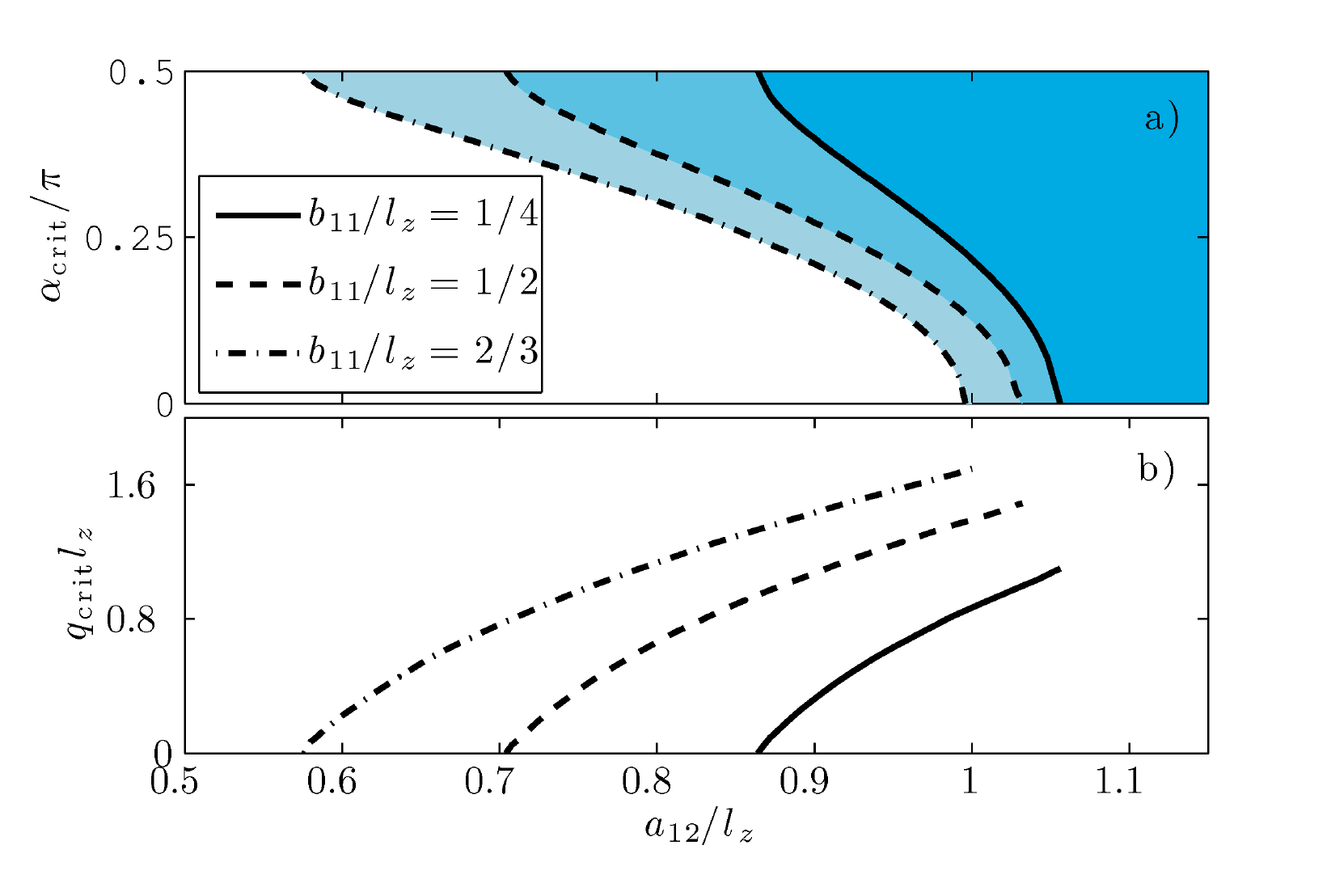}
\caption{\label{fig:paraalpha} (color online) The critical polarization angle $\alpha$ (top panel) and the critical wave number $q_\mathrm{crit} l_z$ (bottom panel) for the balanced two-component q2D DBEC with $a_{11}/l_z = a_{22}/l_z = 1$ and $b_{22}/l_z = 0$ for various dipole lengths of component 1, $b_{11}/l_z$, as a function of the interspecies $s$-wave scattering length $a_{12}/l_z$.  The immiscible parameters are shaded in the top panel.  Here, the roton immiscibility emerges as a function of polarization angle for $\qq_\mathrm{crit}$ perpendicular to the direction of tilt.}
\end{figure}

In figure~\ref{fig:paraalpha}, we extend this result to characterize the onset of roton immiscibility as a function of tilt angle $\alpha$ and interspecies scattering lengths for various $b_{11}/l_z$.  In figure~\ref{fig:paraalpha}(a), the critical scattering length for the transition to immiscibility is shown as a function of the interspecies scattering length for various dipole lengths of component 1, $b_{11}/l_z = 1/4,1/2,2/3$.  Again, we emphasize that component 2 is non-dipolar, so $b_{22}/l_z=0$ here.  In figure~\ref{fig:paraalpha}(b), the corresponding critical wave number $q_\mathrm{crit}l_z$ is shown.  For all dipole lengths, the onset of immiscibility as a function of $a_{12}/l_z$ occurs at $q_\mathrm{crit}l_z=0$ for a critical polarization angle $\alpha_\mathrm{crit} = \pi/2$, corresponding to the dipoles being polarized in the plane of symmetry.  This is expected, though, as the interaction~(\ref{hjk}) exhibits no momentum-dependence for $\alpha=\pi/2$ in the direction perpendicular to the polarization tilt and the criteria for miscibility can thus be extracted from that of the homogeneous\label{sec:mf} two-component DBEC, given in Eq.~(\ref{Deltaq}).  Indeed, the values of $a_{12}/l_z$ for the long-wavelength IMT thresholds at $q_\mathrm{crit}l_z=0$ can be found by setting $\Delta(\pi/2)=0$ and solving for $a_{12}/l_z$.  As $a_{12}/l_z$ is increased beyond these long-wavelength threshold values, however, the critical polarization angle decreases and eventually approaches $\alpha_\mathrm{crit} = 0$, which characterizes the roton immiscibility discussed earlier.  In this parameter range, $q_\mathrm{crit} l_z >0$, and the critical wave number approaches characteristic roton wave numbers for larger $a_{12}/l_z$.

To further demonstrate the role that the polarization direction plays in the roton immiscibility of the two-component q2D DBEC, we model a time-dependent process (perhaps an experimental scenario) in which the polarization field $\alpha$ is tilted as a function of time, driving the system to an immiscible state through manipulating the anisotropy of the ddi in component 1.  We model this scenario via direct time-dependent integration of the coupled GPEs, Eqs.~(\ref{q2DGPEj}).  We take the parameters for the balanced system introduced earlier, with $a_{11}/l_z = a_{22}/l_z = 1$, $a_{12}/l_z = 1.05$, $b_{11}/l_z = 1/4$ and $b_{22}/l_z = 0$.  To begin, we consider $\alpha = 0$ at time $t \omega_z =0$, corresponding to a miscible system.  To break the symmetry of this homogeneous ground state, we seed the condensates with quasiparticles as given in Eqs.~(\ref{imagtimeseed}) and~(\ref{BEdist}), taking $T=100\, \mathrm{nK}$.  Then, we linearly ramp the polarization angle to a final value $\alpha_\mathrm{hold}=\pi/4$ over a time $t_\mathrm{ramp}$, we hold the polarization angle at $\alpha_\mathrm{hold}$ for a time $t_\mathrm{hold}$, and we ramp the polarization angle back to $\alpha=0$ over the time $t_\mathrm{ramp}$.  For this simulation, we take $t_\mathrm{ramp} \omega_z = 200$ and $t_\mathrm{hold} \omega_z = 400$, so that the time scales of the polarization tilt are much greater than the other time scales in the system, being the inverse quasiparticle energies, and the transition is thus to a good approximation adiabatic.  

\begin{figure*}[t!]
\includegraphics[width=2\columnwidth]{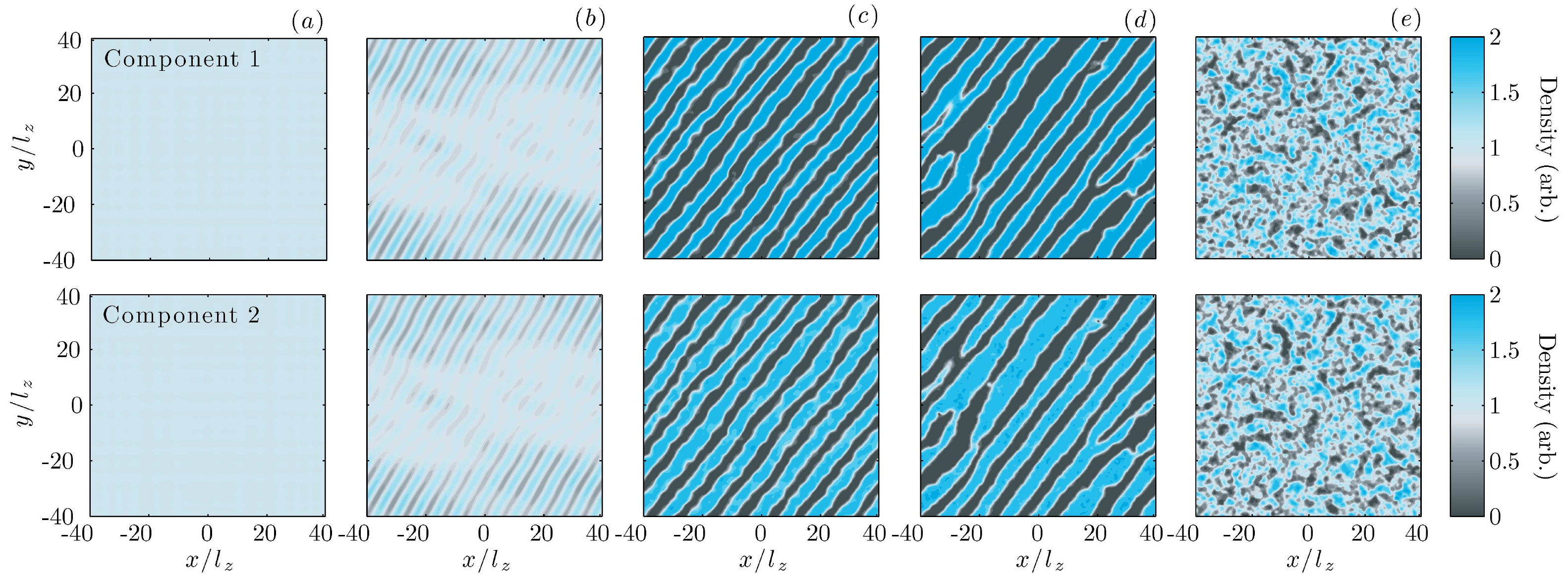}
\caption{\label{fig:tiltsim} (color online) Condensate densities $n_1(\rrho,t) = |\varphi_1(\rrho,t)|^2$ (top row) and $n_2(\rrho,t)=|\varphi_2(\rrho,t)|^2$ (bottom row) as a function of time, where the polarization angle is tilted from a value where the system is miscible ($\alpha=0$ at $t=0$, column (a)) to a value where the system is roton immiscible ($\alpha_\mathrm{hold}=\pi/4$) at an angle $\eta=\pi/5$ off of the $x$-axis.  In (a), the system is clearly miscible.  Over a time $t_\mathrm{ramp}\omega_z=200$, $\alpha$ is linearly ramped from $\alpha=0$ to $\alpha_\mathrm{hold}=\pi/4$.  It is held at $\alpha_\mathrm{hold}$ for $t_\mathrm{hold}\omega_z = 400$, then is linearly ramped back to $\alpha=0$ over $t_\mathrm{ramp}\omega_z=200$.  In column (b), the densities are shown at time $t\omega_z = 120$, where the system is just starting to exhibit immiscible character.  The densities are shown at time $t\omega_z = 200$ in column (c), where the system is fully immiscible on the roton length scale.  The densities are shown for $t\omega_z = 760$ in column (d), as the angle has returned to a value corresponding to miscibility, and column (e) shows the densities at time $t\omega_z = 1200$, where the polarization angle has returned to $\alpha=0$ but they system is not rethermalized due to the introduction of strong phase fluctuations in the process of spatial seperation en route to miscibility.}
\end{figure*}

Because the real-time evolution of the coupled GPEs preserves total energy and is non-dissipative,  we seed the condensate with quasiparticles with wave vectors in the $x$ or $y$ directions only.  This ensures that the periodic boundary conditions of our numeric grid are satisfied and there are no unphysical high-momentum fluctuations at the edges of the system.  As a result, the initial quasiparticles propagate only in the $x$ and $y$ directions.  Thus, choosing a tilt direction of $\eta=0$ or $\eta=\pi/2$ (where $\eta$ is the tilt angle relative to the $x$-axis) would share the symmetry of the initial quasiparticle noise.  Instead, we choose $\eta = \pi/5$.  According to the Bogoliubov theory presented above, the corresponding immiscibility should result in density stripes along the direction $\eta=\pi/5$.

We present results from this simulation in figure~\ref{fig:tiltsim}, where we plot the densities of components 1 and 2 as a function of time.  Column (a) shows the densities at time $t\omega_z=0$, where the densities are approximately equal, aside from the initial seeding that is not visually noticeable in the given contour scaling.  Column (b) shows the densities at time $t\omega_z = 120$, when the polarization angle just exceeds $\alpha_\mathrm{crit}$ and the system begins to exhibit immiscible character.  Column (c) shows the densities at time $t\omega_z = 200$, at the end of the polarization angle ramp, and when the system exhibits full immiscibility.  The immiscibility clearly has roton character, and as anticipated, the direction of the immiscible stripes coincides with the tilt direction, at $\eta=\pi/5$.  Columns (d) and (e) show the densities at times $t\omega_z = 760$ and $t\omega_z = 1200$, respectively.  These columns demonstrate the mechanism for symmetry breaking of the striped roton immiscible state, as the polarization angle tilts back to $\alpha=0$.  While the system remains immiscible, the stripe character is completely lost by $t\omega_z = 1200$.  Here, the introduction of strong phase fluctuations during the transition to immiscibility prevents the system from returning to the miscible state over the time scales considered here.

The issue of thermalization in multicomponent BECs is of increasing interest with the finite temperature studies of $F=1$ spinor condensates, such as the $^{87}$Rb system~\cite{Sadler06,Vengalattore08,Vengalattore10}.  Recently, it was shown that such systems also do not thermalize over even very long time scales, following a quench from the polar to the ferromagnetic state~\cite{Barnett11}.

\subsection{Radially Trapped Case}
\label{sec:radtrap}

Until now, we have considered only the case where the q2D system is homogeneous in the $x$-$y$ plane.  In a realistic experimental scenario, the trapping potential will have a finite radial extent.  For the roton immiscibility to persist in this geometry, we expect that the trapped system must be such that the density of the components is sufficiently large over a transverse length scale that is sufficiently greater than the roton length scale.  To investigate such a claim, we solve the coupled GPEs~(\ref{q2DGPEj}) numerically in the presence of a radial trap with aspect ratio $\lambda=10$, where $\lambda=\omega_z/\omega_\rho$.  While this trap aspect ratio is seemingly small, it serves to model a larger trap aspect ratio, as our solutions are free to extend in the radial direction but are Gaussians with a fixed width in the axial ($z$) direction.

Much like the non-dipolar system, we find that a larger interspecies repulsion (corresponding to a smaller $\Delta$) is necessary for the immiscibility to occur in a radial trap~\cite{Timmermans98}.  Thus, we take $a_{11}/l_z = a_{22}/l_z = 1$ and $a_{12}/l_z = 1.1$ with, as before, $b_{22}/l_z=0$, and investigate solutions for varying $b_{11}/l_z$.  We solve the coupled GPEs via imaginary time propagation, where we take the initial wave functions to be Gaussians with small amplitude random noise sampled on the numeric grid.  We converge the energy of each component to a part in $10^{6}$.

\begin{figure}[t!]
\includegraphics[width=0.95\columnwidth]{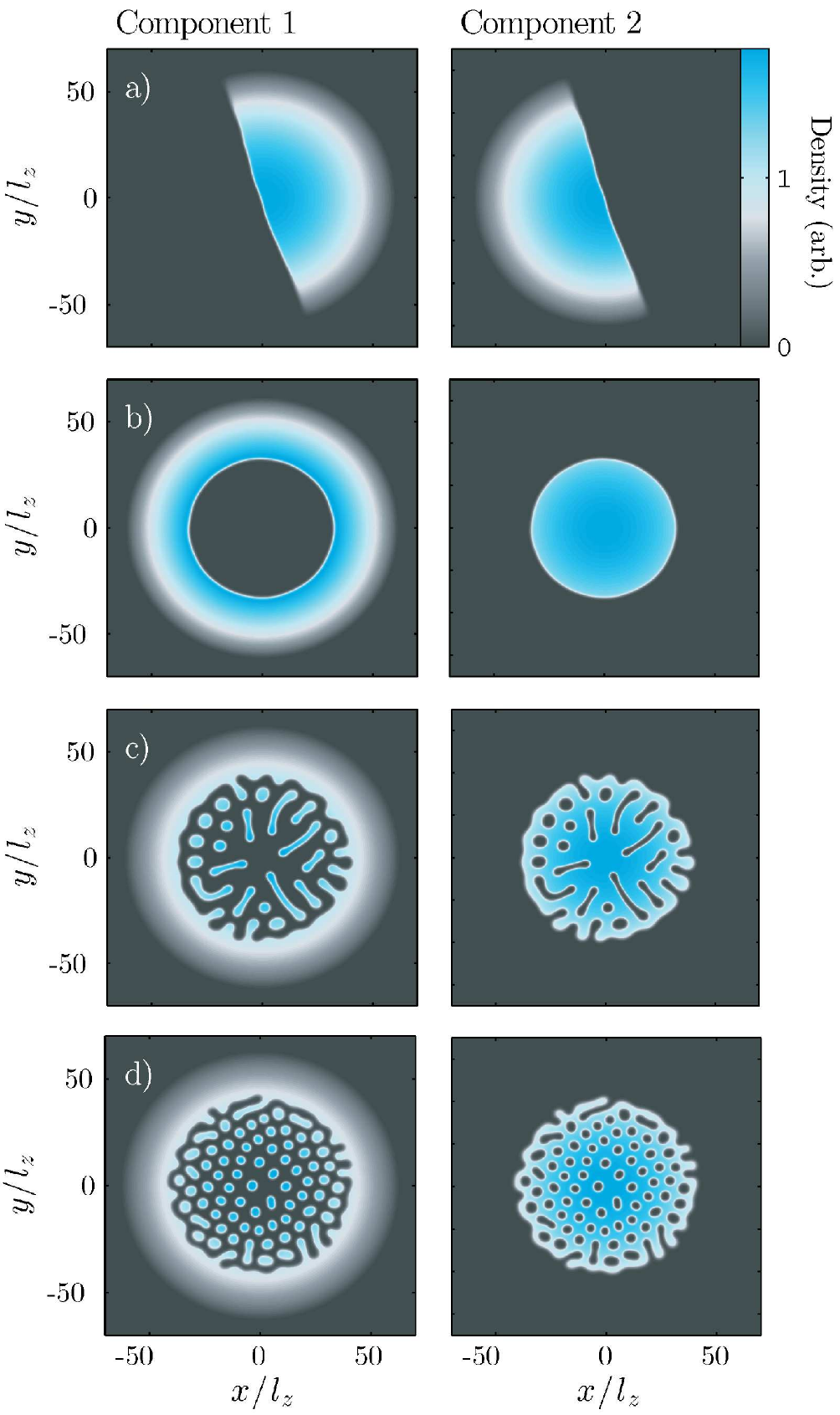}
\caption{\label{fig:trapped} (color online) Stationary densities $n_1(\rrho)=|\varphi_1(\rrho)|^2$ (left column) and $n_2(\rrho)=|\varphi_2(\rrho)|^2$ (right column) of the balanced two-component DBEC in a trap with aspect ratio $\lambda=10$ for $a_{11}/l_z = a_{22}/l_z = 1$, $a_{12}/l_z = 1.1$ and $b_{22}/l_z = 0$.  The rows show solutions for various dipole lengths of component 1, being a) $b_{11}/l_z = 0$, b) $b_{11}/l_z = 1/4$, c) $b_{11}/l_z=1/2$ and d) $b_{11}/l_z = 2/3$.  In row (a), the system in non-dipolar and the interactions are thus balanced, resulting in an immiscibility on the length scale of the trap.  In row (b), the presence of the relatively small dipole moment in component 1 breaks this symmetry, but the immiscibility is still long-wavelength and component 1 thus engulfs component 2 due to its greater self-repulsion.  The roton immiscibility emerges for $b_{11}/l_z \gtrsim 1/2$, as is seen in rows (c) and (d).  While both immiscible states exhibit ordering near the roton length scale, the length scale is smaller in (d) where the dipole length is larger, as suggested by the results shown in figure~\ref{fig:para}. }
\end{figure}

We present some results in figure~\ref{fig:trapped} for (a) $b_{11}/l_z = 0$, (b) $b_{11}/l_z = 1/4$, (c) $b_{11}/l_z = 1/2$ and (d) $b_{11}/l_z = 2/3$.  For $b_{11}/l_z = 0$, the system is non-dipolar and the IMT is characterized by the contact interaction strengths alone.  In this case, the immiscibility is seen as the splitting of the components at the largest available length scale, being the radial extent of the system.  For $b_{11}/l_z = 1/4$, the system is still immiscible, but the ddi is not sufficiently strong to induce roton immiscibility.  Instead, the system exhibits a long wavelength immiscibility, but now the density of component 1 is pushed to the outside of the trap due to the intraspecies interactions of component 1 being greater than those of component 2.  The roton immiscibility emerges near $b_{11}/l_z = 1/2$, as we see in row (c) of figure~\ref{fig:trapped}.  Because we initiate the imaginary time evolution with randomly sampled noise on the numeric grid, the system does not prefer a direction to break the symmetry of the miscible state, as is shown for the homogeneous q2D system in figure~\ref{fig:rotonim}.  Here, the roton immiscibility manifests as the dipolar component forms ``bubbles'' with spacings on the order of $10 l_z$, just larger than the characteristic roton wavelength $\lambda_\mathrm{roton}=2\pi l_z$.  For $b_{11}/l_z = 2/3$, shown in row (d), however, the mean spacing between clumps is on the order of $\lambda_\mathrm{roton}$.

Because of the effectively stronger self-repulsion of component 1 (due to the ddi), it possesses a finite density at radial extents greater than component 2.  This is seen in all cases where $b_{11}/l_z > 0$, in rows (b), (c) and (d) of figure~\ref{fig:trapped}.  Interestingly, for smaller condensate densities, we find that increasing $b_{11}/l_z$ can result in the majority of component 1 being pushed to the outside of the trap, suppressing the roton immiscibility.  For intermediate $b_{11}/l_z$, however, the roton immiscibility persists near the boundary of the two components.  We find that such a phenomenon exists in more oblate traps, as well, where the radial trapping potential does not force the components to overlap in high density regions.

It is interesting to note that the results in figure~\ref{fig:trapped} are very similar to those presented in~\cite{Saito09}, where a dipolar component and a non-dipolar component are separated in the direction of polarization by an external magnetic field gradient, forcing immiscibility in this direction.  For a sampling of interaction strengths and field gradients, pattern formations on finite length scales, reminiscent of a classical magnetic ferrofluid, are predicted by direct solutions to the coupled GPEs that describe the system.  These patterns are similar to those seen in figure~\ref{fig:trapped}, though perhaps from a different physical origin, and depend strongly on noise that is used to break the symmetry of the initial guess for the condensate wave functions.

As is clear from our preceding discussion, the roton immiscibility depends strongly on the interaction parameters in the system under consideration, and special care must be taken in proposing a candidate species with which to investigate such a phenomenon.  One such candidate is the ground $^7 S_3$ state of atomic $^{52}$Cr prepared in the $m_J=0$ spin projection for the non-dipolar component and the $m_J=-3$ spin projection for the dipolar component, which has a  magnetic dipole moment of $d=6 \mu_B$ where $\mu_B$ is the Bohr magneton.  While the relaxation lifetimes of the $m_J>-3$ projections of the $^7 S_3$ state of $^{52}$Cr were measured to be relatively long~\cite{Griesmaier05}, spin-exchange collisions, occurring on a time scale $\sim .1\,\mathrm{s}$, may limit the experimental feasibility of using $^{52}$Cr.  Nevertheless, for a trap with an axial frequency of $\omega_z = 2\pi \times 2\,\mathrm{kHz}$, the lifetime of the roton immiscibility, for the maximum imaginary part of $\omega_-(\qq)$ being $\sim 0.1 \omega_z$,  is characteristically $\sim 0.80 \, \mathrm{ms}$.  The critical integrated density for roton immiscibility in this case is $n_\mathrm{2D} \sim 4.0\times 10^{11}\,\mathrm{cm}^{-2}$, corresponding to a maximum 3D density of $n_\mathrm{3D} \sim 7.2 \times 10^{15} \, \mathrm{cm}^{-3}$.  The immiscibility for the case seen in figure~\ref{fig:trapped}(c) with trap with axial frequency $\omega_z = 2\pi \times 2\,\mathrm{kHz}$ would require a total number of $^{52}$Cr atoms on the order $N\sim 6.2 \times 10^6$ with scattering lengths $a_{11}= a_{22} = 30.4\, a_0$ and $a_{12} = 33.4\, a_0$, where $a_0$ is the Bohr radius.  This is a case that is possibly achievable by sufficient manipulation of the Fano-Feshbach resonances~\cite{Werner05}.

For more strongly interacting species, however, the roton immiscibility emerges for a smaller critical density or particle number.  Other possible dipolar species include atomic Dy, which has been recently Bose-condensed~\cite{Lu11PRL}, and polar molecules, which can possess relatively large electric dipole moments.  For the trap discussed above, the roton immiscibility emerges for a total number of $^{164}$Dy atoms of $N\sim 400 \times 10^3$, or a critical maximum density of $n_\mathrm{3D} \sim 2.6 \times 10^{15}\, \mathrm{cm}^{-3}$.  An experimental study of the magnetic Fano-Feshbach resonances in atomic Dy, however, has yet to be performed.  For polar molecules of, say, RbCs~\cite{Debatin11arXiv} with an electric dipole moment of $d=0.5\,\mathrm{Db}$, a critical density of $n_\mathrm{3D}\sim 9.0 \times 10^{13}\,\mathrm{cm}^{-3}$, or a critical particle number of $N\sim 8900$ molecules is needed.  The realization of one dipolar and one non-dipolar component for the molecular case is unclear, however, a mixture of, say, RbCs and Rb may be possible in the near future.  Additionally, we have checked that roton immiscibility exists for an appropriate set of $s$-wave scattering lengths in alkali atom (with $\sim 1\,\mu_B$ magnetic dipole moments) and Cr or Dy mixtures.

\section{Conclusion}
\label{sec:conc}

The long-range and anisotropic nature of the ddi plays an interesting, nontrivial role in the physics of many-body systems.  Here, we focus on the case of a two-component Bose-Einstein condensate and show that, for a set of specific interaction parameters and trap geometries, the system exhibits immiscibility with roton character, where the immiscibility in a non-dipolar system or a homogeneous 3D dipolar system is strictly phonon-like.  In particular, we find that the roton immiscibility occurs in the q2D geometry when the interspecies and intraspecies $s$-wave scattering lengths have comparable values, while one component is not (or negligibly) dipolar and the other component possesses a dipolar length that is comparable to the $s$-wave scattering lengths.  By employing the Bogoliubov theory to the q2D homogeneous two-component system, we calculated a two-component dispersion to efficiently characterize the stability of the miscible state and the parameters that define the IMT threshold for the familiar phonon-like and the roton immiscibilities.  Direct simulations of the coupled GPEs verify these results and reveal interesting dynamic and symmetry-breaking features of the immiscible phase.  Recent experimental progress inspires confidence that the roton immiscibility may be observable in the near future.

\section{Acknowledgments}
\label{sec:ack}

RMW  and JLB acknowledge financial support from the NSF.  CT acknowledges support from the Advanced Simulation and Computing Program (ASC).  CT and ET acknowledge support from LANL which is operated by LANS, LLC for the NNSA of the U.S. DOE under Contract No. DE-AC52- 06NA25396.

\end{document}